\documentclass[aps,prd,showpacs,twocolumn,floatfix]{revtex4}
\usepackage{epsfig,dcolumn}
\usepackage{graphicx}
\usepackage{bm}
\newcommand{\nc}{\newcommand}
\nc{\mb}[1]{\makebox[#1]{}}
\nc{\CC}{{\scriptscriptstyle CC}}
\nc{\NC}{{\scriptscriptstyle NC}}
\nc{\V}{{\rm v}}
\nc{\W}{{\scriptscriptstyle W}}
\nc{\X}{{\scriptscriptstyle X}}
\nc{\Z}{{\scriptscriptstyle Z}}
\nc{\CS}{{\scriptscriptstyle CS}}
\nc{\DY}{{\scriptscriptstyle DY}}
\nc{\PW}{{\scriptscriptstyle PW}}
\nc{\SB}{{\scriptscriptstyle SB}}
\nc{\CSV}{{\scriptscriptstyle CSV}}
\nc{\GLS}{{\scriptscriptstyle GLS}}
\nc{\CIB}{{\scriptscriptstyle CIB}}
\nc{\PT}{{\scriptscriptstyle PT}}
\nc{\ASYM}{{\scriptscriptstyle asym}}
\nc{\IE}{{\it i.e.,\ }}
\nc{\EG}{{\it e.g.,\ }}
\nc{\EA}{{\it et al.,\ }}
\nc{\AH}{{\it ad hoc\ }}
\nc{\CHPT}{{$\chi_{\PT}$\ }}

\nc{\NCA}{{\em Nuovo Cimento}}
\nc{\NIM}{{\em Nucl. Instrum. Methods}}
\nc{\NIMA}{{\em Nucl. Instrum. Methods} A}
\nc{\NPB}{{\em Nucl. Phys.} B}
\nc{\PLB}{{\em Phys. Lett.}  B}
\nc{\PRL}{{\em Phys. Rev. Lett.}}

\nc{\PRD}{{\em Phys. Rev.} D}
\nc{\PRC}{{\em Phys. Rev.} C}
\nc{\ZPC}{{\em Z. Phys.} C}

\nc{\st}{\scriptstyle}
\nc{\sst}{\scriptscriptstyle}
\nc{\mco}{\multicolumn}
\nc{\epp}{\epsilon^{\prime}}
\nc{\vep}{\varepsilon}
\nc{\ra}{\rightarrow}
\nc{\ppg}{\pi^+\pi^-\gamma}
\nc{\nuN}{{\nu N_0}}
\nc{\nubN}{{\overline{\nu} N_0}}
\nc{\ovnu}{{\overline{\nu}}}
\nc{\snuNC}{{\langle \sigma^{\nuN}_{\NC}\rangle }}
\nc{\snubNC}{{\langle \sigma^{\nubN}_{\NC}\rangle }}
\nc{\snuCC}{{\langle \sigma^{\nuN}_{\CC}\rangle }}
\nc{\snubCC}{{\langle \sigma^{\nubN}_{\CC}\rangle }}
\nc{\Rnu}{{R^{\nu}}}
\nc{\Rnub}{{R^{\overline{\nu}}}}
\nc{\sintW}{{\sin^2 \theta_{\W} }}
\nc{\MS}{{\overline{MS}}}
\nc{\vp}{{\bf p}}
\nc{\rz}{{\rho_0^2}}
\nc{\ko}{K^0}
\nc{\kb}{\bar{K^0}}
\nc{\al}{\alpha}
\nc{\ab}{\bar{\alpha}}
\nc{\be}{\begin{equation}}
\nc{\ee}{\end{equation}}
\nc{\bea}{\begin{eqnarray}}
\nc{\eea}{\end{eqnarray}}
\nc{\beast}{\begin{eqnarray*}}
\nc{\eeast}{\end{eqnarray*}}

\begin{document}

\titlepage

\title{Constraints on Parton Charge Symmetry and Implications for 
Neutrino Reactions} 

\author{J.T.Londergan}

\email{tlonderg@indiana.edu}
\affiliation{Department of Physics and Nuclear
            Theory Center,\\ Indiana University,\\ 
            Bloomington, IN 47405, USA}

\author{A.W.Thomas}
\email{awthomas@jlab.org}
\affiliation {Jefferson Lab, 12000 Jefferson Ave.,\\ 
                Newport News, VA 23606, USA}
\date{\today}


\begin{abstract} 
{}For the first time, charge symmetry breaking terms in parton 
distribution functions have been included in a global fit to high energy 
data.  We review the results obtained for both valence and sea quark 
charge symmetry violation, and we compare these results with the most 
stringent experimental upper limits on charge symmetry violation for 
parton distribution functions, and with theoretical estimates of 
charge symmetry violation. The limits allowed in the global fit would 
tolerate a rather large violation of charge symmetry. We discuss the 
implications of this for the extraction of the Weinberg angle in 
neutrino DIS by the NuTeV collaboration. 
\end{abstract}




\maketitle


\section{Introduction\label{Sec:Intro}}

Charge symmetry is a restricted form of isospin 
invariance involving a rotation of $180^\circ$ about the ``2'' axis 
in isospin space.  For parton distributions, charge symmetry involves 
interchanging up and down quarks while simultaneously interchanging 
protons and neutrons.  In nuclear physics, charge symmetry is generally 
obeyed at the level of a fraction of a percent~\cite{Miller,Henley}.  
Charge symmetry violation in parton distribution functions (PDFs) arises 
from two sources; 
from the difference $\delta m \equiv m_d - m_u$ between down and up current 
quark masses, and from electromagnetic (EM) effects.  

Since charge symmetry is so well satisfied at lower energies, it is 
natural to assume that it holds for parton distributions.  
Introducing charge symmetry reduces by a factor of two the number 
of PDFs necessary to describe high-energy data.  In addition, as we 
shall see, there is no direct experimental evidence that points to a 
substantial violation of charge symmetry in parton distributions.  
For this reason, until recently all phenomenological PDFs have 
assumed charge symmetry at the outset. In Sec.\ \ref{Sec:Csymmexp}, 
we review the experimental evidence for charge symmetry in PDFs.  
Recent experiments allow us to place reasonably strong upper 
limits on parton charge symmetry violation.  In Sec.\ \ref{Sec:Csymmth}, 
we review some theoretical estimates for parton charge symmetry 
violation (CSV) for valence quarks, and we compare this with the 
experimental 
limits.  In Sec.\ \ref{Sec:Csymmph}, we review the recent 
global fit by Martin \EA \cite{MRST03} that includes for the first 
time the possibility of charge symmetry violating PDFs both for 
valence and for sea quarks.  We compare the results of this 
phenomenological fit with the experimental limits on parton 
CSV as well as with theoretical estimates of both the magnitude and 
sign of charge symmetry violating parton distributions.   

In Sect.\ \ref{Sec:CSVPW}, we review the effect 
of isospin violating PDFs on neutrino DIS measurements.  In particular, 
we concentrate on the extraction of the Weinberg angle by the 
NuTeV collaboration \cite{NuTeV}.  We show the magnitude of 
the effects predicted by theoretical calculations, and we  
contrast this with the magnitude of effects allowed by the 
phenomenological CSV PDFs extracted by the MRST group \cite{MRST03}.  

\section{Experimental Limits on Parton Charge 
Symmetry Violation \label{Sec:Csymmexp}}

There are no direct measurements that reveal the presence of charge 
symmetry violation 
in parton distribution functions.  At present, we have only upper 
limits on the magnitude of charge symmetry violation.  

The most stringent test of parton charge symmetry comes from comparing  
the structure function $F_2^{\nu}$ measured in neutrino induced charged 
current reactions, and the structure function $F_2^{\gamma}$ for charged 
lepton DIS, both measured on isoscalar targets $N_0$.  In leading order, 
assuming parton charge symmetry, the structure functions have the form 
\cite{Lon98} 
\bea
F_2^{\gamma N_0}(x)  &\approx& {5 Q(x)\over 18}\,  
 + {x\over 6} \left(c(x) + \bar{c}(x) -s(x) - \bar{s}(x)\right)  \nonumber \\ 
 F_2^{W^{\pm} N_0}(x) &\approx& Q(x) \pm x\left( s(x) - \bar{s}(x) + 
 \bar{c}(x) - c(x) \right) \nonumber \\ 
 Q(x) &=& \sum_{j=u,d,s,c} x\left( q_j(x) + \bar{q}_j(x) \right)  
\label{eq:F2gamm}
\eea
In Eq.\ \ref{eq:F2gamm}, $F_2^{W^{\pm} N_0}$ is the structure function for 
charged-current processes induced by neutrinos (antineutrinos) on an 
isoscalar target.  Charge symmetry violation and NLO effects 
are not included in this equation.  In the limit of exact 
charge symmetry, there is a simple relation between the charged-lepton and 
neutrino structure functions, corrected for heavy quark effects.  This 
relation is defined as the ``charge ratio'' $R_c(x,Q^2)$ or, as it is 
sometimes termed, the ``5/18$^{th}$ rule.''  The factor $5/18$ is simply 
understood as the average of the squares of the light quark charges, 
relative to the weak charges.  Expanding $R_c$ to lowest order in the 
(presumably small) charge symmetry violating terms gives  
\bea
  R_c(x) &\equiv&  
 {F_2^{\gamma N_0}(x) +x\left( s(x) + \bar{s}(x) -c(x) - \bar{c}(x) 
 \right)/6 \over 5\overline{F_2}^{W N_0}(x)/18 } \nonumber \\ 
 &\approx& 1 + {3x \left( \delta u(x) + \delta \bar{u}(x) - \delta d(x) 
 - \delta \bar{d}(x)\right) \over 10 \,Q(x) } \nonumber \\ 
\label{eq:Rc}
\eea
Eq.\ \ref{eq:Rc} introduces the CSV parton distributions, 
\bea
 \delta u(x) &=& u^p(x) - d^n(x) \, ; \nonumber \\  
 \delta d(x) &=& d^p(x) - u^n(x) \, , 
\label{eq:CSVdef}
\eea 
with analogous relations for antiquarks.  The quantity 
$R_c(x)$ in Eq.\ \ref{eq:Rc} is defined using 
$\overline{F_2}^{W N_0}$, the average of neutrino and antineutrino,  
$F_2$, charge-changing structure functions; it also requires knowledge 
of strange and charm PDFs.  Deviation of $R_c(x)$ from unity would be an 
indication of a non-zero charge symmetry violating contribution.   

The most precise neutrino measurements were obtained by the 
CCFR group \cite{Sel97}, who extracted the $F_2$ structure
function for neutrino and antineutrino interactions on iron using
the Quadrupole Triplet Beam at FNAL.  This can be compared with several 
measurements of the $F_2$ structure functions from DIS reactions using 
high-energy muons or electrons.  The most precise measurements were 
obtained by the NMC group \cite{Ama91,Arn97}, who measured $F_2$ structure 
functions for muon interactions on deuterium at muon energies $E_\mu = 90$ 
and 280 GeV. Earlier measurements were obtained by the BCDMS muon 
scattering experiments on deuterium \cite{Ben90} and carbon \cite{BCDMS}, 
and electron scattering results from SLAC \cite{Whi90,Whi90b}.

Precision measurements of the charge ratio require a significant number of 
corrections.  It is necessary to know the relative normalization between 
charged-lepton and neutrino cross sections. Heavy quark threshold effects 
are very important, as the most important neutrino energies are sufficiently 
small that one must correct for the finite mass 
of the charmed quark~\cite{Steffens:1999hx}.  
The CCFR neutrino cross sections were taken on iron targets, so one must 
correct for nuclear effects (Fermi motion at large $x$, EMC effects at 
intermediate $x$ and shadowing at small $x$ 
\cite{Arn84,Gom94,Ama95,Ada95,Selth,Miller:2002xh}).  There are additional 
corrections since iron is not an isoscalar target.  Eq.\ \ref{eq:Rc} 
also requires contributions from strange and charmed quarks.  The CCFR  
\cite{Baz93} and NuTeV \cite{Gon01} groups have extracted strange 
(antistrange) quark distributions from 
the cross sections for opposite sign dimuons in reactions induced by 
neutrinos (antineutrinos).  

\begin{figure}
\includegraphics[width=3.0in]{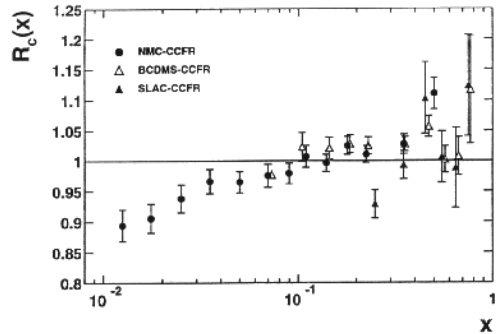}
\caption{The charge ratio of Eq.\ \protect{\ref{eq:Rc}} vs.\ $x$ 
using CCFR neutrino data of Ref.\ \protect{\cite{Sel97}} combined  
with muon DIS data. Solid circles: NMC $\mu-D$ data, 
Ref.\ \protect{\cite{Ama91}}; open triangles: muon measurements from the BCDMS 
group on $D$, Ref.\ \protect{\cite{Ben90}} and $C$, 
Ref.\ \protect{\cite{BCDMS}}; solid triangles: SLAC electron 
scattering data Refs.\ \protect{\cite{Whi90,Whi90b}}.
\label{Fig:fig45a}}
\end{figure}

Fig.\ \ref{Fig:fig45a} plots the 
charge ratio $R_c$ of Eq.\ \ref{eq:Rc} vs.\ $x$.  The circles are
the NMC/CCFR ratio.  The open triangles are the BCDMS/CCFR charge
ratio, where BCDMS represents the muon scattering results of the
BCDMS group on deuterium \cite{Ben90} and carbon \cite{BCDMS}.  The
solid triangles are the SLAC/CCFR charge ratio, where SLAC denotes 
electron scattering results of the SLAC group \cite{Whi90,Whi90b}. 
Since the CCFR measurements were obtained from neutrino-iron 
scattering, and the NMC measurements were from $\mu-D$ reactions, the 
NMC measurements were ``converted'' to the equivalent reactions on 
iron by multiplying by the ratio of $\mu-Fe$ to $\mu-D$ cross 
sections \cite{Shadnt,Bor98b}.  In addition, a correction for strange 
quarks and charm quarks was applied to the numerator of Eq.\ 
\ref{eq:Rc}.  

In the region $0.1 \le x \le 0.4$ the charge ratio test is 
consistent with unity, with errors in the range $2-3$\%.  Solving 
Eq.\ \ref{eq:Rc} for the CSV parton distributions, 
\be
 {x\left( \delta u(x) + \delta \bar{u}(x) - \delta d(x) 
 - \delta \bar{d}(x)\right) \over Q(x) } = {10 \over 3}\left( R_c - 1 \right) 
\label{eq:partCSV}
\ee
one can set an upper limit 
to parton CSV effects in this $x$ range at about the $6-9$\% level.  
For larger values of $x$ the upper limit on CSV effects is substantially 
greater.  This is due both to poorer statistics and to the 
large Fermi motion corrections needed for the heavy 
target at large $x$.   

At values $x < 0.1$, the charge ratio appeared to deviate significantly 
from unity.  The discrepancy increased monotonically as $x$ decreased, 
and approached 15\% at the smallest values of $x$ 
\cite{Shad}.  Several suggestions 
were advanced to explain this discrepancy, including anomalously large  
contributions from strange quarks \cite{Bro96}, and surprisingly large 
contributions from sea quark charge symmetry violation \cite{Bor98}. 
Eventually the CCFR group re-analyzed the neutrino data, and after 
the new analysis the small-$x$ discrepancy disappeared \cite{Yan02}, 
largely as a result of two factors.  The first was the treatment 
of charm mass corrections.  In the initial analysis, these were 
incorporated using the ``slow rescaling'' hypothesis due to Georgi 
and Politzer \cite{Geo76}.  The re-analysis involved NLO calculations, which  
differed significantly from the slow-rescaling procedure at 
small $x$  -- as had been suggested by Boros \textit{et al.} \cite{Bor00}. 
Since the NLO effects depend on both $x$ and $Q^2$, the   
final results cannot be plotted simply versus Bjorken $x$. 

The second significant effect involved the separation of the $F_2$ 
and $F_3$ structure functions in charged-current neutrino DIS.  The 
sum of neutrino and antineutrino charged-current DIS cross sections 
contains a linear combination of 
neutrino $F_2$ and $F_3$ structure functions, 
\bea 
{d^2 \sigma^{\nu}_{\CC} \over dx dy} + 
 {d^2 \sigma^{\bar{\nu}}_{\CC} \over dx dy} &\sim& 
 2(1- y - y^2/2) \overline{F}_2(x,Q^2) \nonumber \\ 
  &+& (y- y^2/2)\Delta xF_3(x,Q^2) \ . 
\label{eq:sigmaCC} 
\eea
In Eq.\ \ref{eq:sigmaCC}, $\overline{F}_2$ is the average of the 
$F_2$ structure functions for neutrinos and antineutrinos, and in 
leading order (assuming charge symmetry), $\Delta xF_3 = 2x(s + 
\bar{s} - c - \bar{c})$.  The structure functions are 
multiplied by coefficients 
that depend on the invariant $y = p\cdot q/p \cdot k$, where $k (p)$ is 
the four momentum of the initial lepton (nucleon), and $q$ is the 
four momentum of the virtual $W$ exchanged in the interaction.  
In Eq.\ \ref{eq:sigmaCC} we have dropped terms of order 
$m_N^2/s$, and for simplicity the longitudinal/transverse ratio $R$ has 
been set to zero. 

In the initial data analysis, the data for a given $x$ bin was 
averaged over all $y$, and the $\Delta xF_3$ structure function was estimated 
using phenomenological PDFs.  The re-analysis included the $x, y$ 
and $Q^2$ dependence of the cross sections.  
In this way the group was able to extract both $\overline{F}_2$ and 
$\Delta xF_3$.  The experimental values for $\Delta xF_3$ differed 
substantially from the phenomenological predictions.  This affected the 
extracted values for the $F_2$ neutrino structure functions.  The 
description of charm production also plays a significant role in 
determining the value of $\Delta xF_3$ \cite{Kra96,Tho98,Aiv94}.  The 
combined effect of the NLO treatment of charm production, and 
the model-independent extraction of $\Delta xF_3$ removed the small-$x$ 
discrepancy.  The charge ratio $R_c$ of Eq.\ \ref{eq:Rc} 
is now unity to within experimental errors, even at 
small $x$.  Since the NLO treatment depends separately on $Q^2$ and 
$x$, it is no longer possible to display the results in a $Q^2$-independent 
plot such as Fig.\ \ref{Fig:fig45a}.    

\section{Theoretical Estimates of Parton Charge Symmetry Violation 
\label{Sec:Csymmth}}

Both the quark mass difference and EM effects responsible for parton 
charge symmetry violation represent small 
changes to quark PDFs.  Several theoretical estimates of 
parton CSV for valence quarks can be obtained by examining simple models for 
parton distribution functions, then observing how these change upon applying 
the operations of charge symmetry.  

The Adelaide group 
\cite{Signal:yc,Schreiber:1991tc,Adl90,Lon98} 
developed a method for calculating twist-two valence parton distributions 
through the relation 
\be 
q_{\V}(x, \mu^2) = M\, \sum_X \, |\langle X | \psi_+(0) | N\rangle |^2 
 \delta( M(1-x) - p_{\X}^+ ) \, .  
\label{eq:qvAdl}
\ee
In Eq.\ \ref{eq:qvAdl}, $\psi_+ = (1+ \alpha_3)\psi/2$ is the 
operator that removes a quark or adds an antiquark to 
the nucleon state $|N\rangle$, $\mu^2$ is the starting 
scale (where QCD is best approximated by a valence dominated quark model) for 
the quark distribution, $| X\rangle$ represents all 
possible final states that can be reached with this operator   
(\IE $| X\rangle = 2q, 3q + \overline{q}, 4q + 2\overline{q}, \dots$),  
and $p_{\X}^+$ is the plus component ($p^+ \equiv p_3 + E(p)$) of the 
momentum of the residual system. We note that the advantage of using Eq.~\ref{eq:qvAdl}
is that the correct support for the PDF is assured, regardless of the approximation used to evaluate the nonperturbative matrix element $ \langle X | \psi_+(0) | N\rangle $.  

Sather \cite{Sat92} assumed that the dominant result of 
electromagnetic effects was manifested in the neutron-proton mass 
difference $\delta M \equiv M_n - M_p$. At the low $Q^2$ appropriate to 
quark models, the largest contribution to valence PDFs at large $x$ 
arises from configurations where one quark is removed from  
three valence quarks, leaving a residual diquark.  Since CSV 
corresponds to  interchanging up and down quark, and neutron-proton labels, 
valence quark CSV is obtained by examining how Eq.\ \ref{eq:qvAdl} 
changes with $\delta m$ and $\delta M$.  

Restricting Eq.\ \ref{eq:qvAdl} to the residual diquark state $X=2$ 
(valid for large $x$) and neglecting the dependence on transverse momentum 
in the $\delta$-function, Sather obtained analytic relations between 
valence quark CSV and derivatives of the valence PDFs,   
\bea 
 \delta d_{\V}(x) &\equiv& d_{\V}^p(x) - 
  u_{\V}^n(x) \nonumber \\ 
  &=& -\frac{\delta M}{M} \frac{d}{dx} \left[ 
 x d_{\V}(x)\right] - \frac{\delta m}{M} \frac{d}{dx} d_{\V}(x) \nonumber \\
 \delta u_{\V}(x) &\equiv& u_{\V}^p(x) - d_{\V}^n(x) \nonumber \\ 
  &=& \frac{\delta M}{M} \left( - \frac{d}{dx}\left[ 
 x u_{\V}(x)\right] + \frac{d}{dx} u_{\V}(x) \right) 
\label{eq:Satanl}
\eea   
In Eq.\ (\ref{eq:Satanl}), $M$ is the average nucleon mass, $\delta M = 1.3$ 
MeV is the n-p mass difference, and $\delta m = m_d - m_u \sim 
4$ MeV is the down-up quark mass difference.  
From Eq.\ \ref{eq:Satanl}, at large $x$ the down quark 
valence distribution in the proton is expected to be larger than the 
up quark distribution in the neutron; similarly, the down quark 
distribution in the neutron is larger than the up quark distribution 
in the proton at large $x$.  The lowest moment of the valence quark 
distributions is fixed by quark normalization, since  
\bea
 \int_0^1 \,dx \delta d_{\V}(x) &=& \int_0^1 \,dx \delta u_{\V}(x) 
  = 0 \, . 
\label{eq:quarknorm} 
\eea  
A violation of Eq.\ \ref{eq:quarknorm} would be equivalent to 
changing the total number of valence quarks in the proton and/or neutron. 
Eq.\ \ref{eq:quarknorm} shows that if $\delta d_{\V}(x)$ is 
positive at large $x$, it must therefore be negative at small $x$ 
such that the integral over all $x$ vanishes; similarly, the quantity 
$\delta u_{\V}(x)$ must also change sign at small $x$.     

\begin{figure}
\includegraphics[width=4.0in]{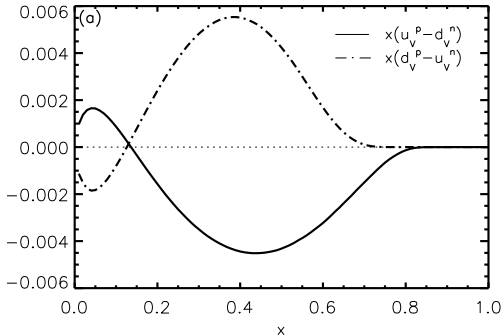}
\caption{Valence quark CSV contributions, $x\delta q_{\V}(x)$ 
vs.\ $x$. Solid line: $x\delta u_{\V}$; dash-dot line: $x\delta d_{\V}$.  
Calculated by Rodionov \EA Ref.\ \protect{\cite{Rodionov:cg}} using MIT 
bag model wavefunctions, evolved to $Q^2 = 10$ GeV$^2$.
\label{Fig:Rodionov}}
\end{figure}

Sather's results do not include effects from configurations with 
more than two quarks in the final state.  In addition, his analytic 
results include only the contribution from quark longitudinal momentum 
(effects from quark transverse momentum in Eq.\ \ref{eq:qvAdl} are 
neglected).  Rodionov \EA \cite{Rodionov:cg} explicitly included the effect  
quark transverse momentum within the MIT bag model.  They did not use the 
approximate equations of Sather, but included quark and nucleon mass 
differences directly in Eq.\ \ref{eq:qvAdl}.  Their results for 
valence quark CSV PDFs are shown in Fig.\ \ref{Fig:Rodionov}.  The solid 
curve is $x \delta u_{\V}(x)$, while the dot-dashed curve is 
$x \delta d_{\V}(x)$, both evolved to $Q^2 = 10$ GeV$^2$.   
Qualitatively, the results of Rodionov \textit{et al.}  are very similar 
to Sather's.  The sign and relative magnitude of both $\delta d_{\V}(x)$ 
and $\delta u_{\V}(x)$ are quite similar in both calculations, and 
the second moment of both distributions is equal to better than  
10\%.  

\section{Phenomenological Charge Symmetry 
Violating PDFs\label{Sec:Csymmph}}

Because CSV effects are typically very small at nuclear 
physics energy scales \cite{Miller,Henley}, from the lack of direct 
evidence for violation of parton charge symmetry, and because 
theoretical estimates put parton CSV at below the 1\% level 
\cite{Lon98}, all previous phenomenological parton distribution functions 
have assumed the validity of parton charge symmetry at the outset.  
However, Martin, Roberts, Stirling and Thorne (MRST) \cite{MRST03} have 
recently studied the uncertainties in parton distributions arising 
from a number of factors, including isospin violation.  

The MRST group chose a specific model for valence quark charge 
symmetry violating PDFs.  They constructed a function that automatically 
satisfied the quark normalization condition of Eq.\ \ref{eq:quarknorm}, 
namely:
\bea 
 \delta u_{\V}(x) &=& - \delta d_{\V}(x) = \kappa f(x) \nonumber \\ 
  f(x) &=& (1-x)^4 x^{-0.5}\, (x - .0909)  \, .
\label{eq:CSVmrst}
\eea
The function $f(x)$ was chosen so that at both small and large $x, f(x)$ 
has the same form as the MRST valence quark distributions, and the 
first moment of $f(x)$ is zero. The functional form of the valence 
CSV distributions guaranteed that $\delta u_{\V}$ and $\delta d_{\V}$ would 
have opposite signs at large $x$, in agreement with the theoretical results 
shown in Fig.\ \ref{Fig:Rodionov}.  

Inclusion of a valence quark CSV term changes the up and down quark 
distributions in the neutron from their charge symmetric counterparts 
in the proton. This changes the momentum carried by  
valence quarks in the neutron from those in the proton, since the 
total momentum carried by valence quarks is given by the second 
moment of the distribution, \EG the momentum carried by up valence 
quarks in the neutron 
\beast
 U^n_{\V} &\equiv& \int_0^1 x\,u^n_{\V}(x) \, dx \ . 
\eeast     
Because the total momentum carried by valence (up plus down) quarks in the 
neutron, $U^n_{\V} + D^n_{\V}$, is rather precisely determined, MRST 
chose a functional form that kept this quantity relatively constant.  
For this reason, they insisted that the valence CSV terms for up 
and down quarks in the nucleon be equal and opposite.  Note that QCD 
evolution does not preserve this relation, but it is very nearly 
maintained for the region of evolution of interest.  The overall coefficient,  
$\kappa$, was then varied in a global fit to a wide range of high 
energy data.  

\begin{figure}
\includegraphics[width=2.2in]{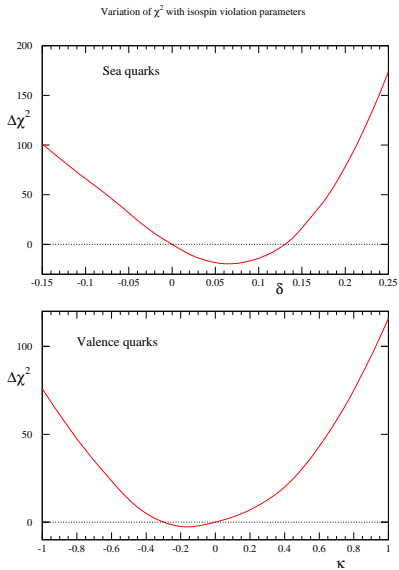}
\caption{$\chi^2$ profile for phenomenological isospin violating 
parton distributions, for sea quarks (top curve) and valence quarks 
(bottom curve), from the MRST 
group, Ref.\ \protect{\cite{MRST03}}. The quantity $\delta$ associated 
with sea quark isospin violation is defined  in Eq.\ 
\protect{\ref{eq:seaCSV}}, while the coefficient $\kappa$ is defined in 
Eq.\ \protect{\ref{eq:CSVmrst}}.
\label{Fig:chi2}}
\end{figure}

\begin{figure}
\includegraphics[width=2.2in]{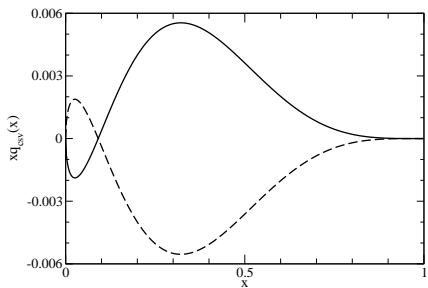}
\caption{The valence quark CSV function from Ref.\ \protect{\cite{MRST03}}, 
corresponding to best fit value $\kappa = -0.2$ defined in Eq.\ 
\protect\ref{eq:CSVmrst}. Solid curve: 
$x\delta d_{\V}(x)$; dashed curve: $x\delta u_{\V}(x)$.
\label{Fig:MRSTfx}}
\end{figure}

The value of $\kappa$ which minimised $\chi^2$ was $\kappa = -0.2$.  The 
MRST $\chi^2$ vs.\ $\kappa$ is shown as the bottom curve in 
Fig.\ \ref{Fig:chi2}.  
Clearly $\chi^2$ has a shallow minimum with the 90\% confidence level 
obtained for the range $-0.8 \le \kappa \le +0.65$.  Since the 
form chosen by MRST for valence quark CSV is strongly constrained, 
and the resulting $\chi^2$ minimum is quite shallow, one should not  
assign too much significance to their result. In global fits of 
this type, one should remember that, unless the shape of the constrained 
function is in close agreement with the actual parton distribution, 
the overall magnitude obtained in a global fit can be misleading. In Fig.\ 
\ref{Fig:MRSTfx} we plot the valence quark CSV PDFs corresponding to the 
MRST best fit value $\kappa = -0.2$.  They  look extremely similar to the 
theoretical valence 
quark PDFs calculated by Rodionov {\it et al.} and shown in Fig.\  
\ref{Fig:Rodionov}; they are also in good agreement with Sather's 
valence CSV distributions.  This provides some theoretical 
support for the functional form chosen by MRST.  However, within the 
90\% confidence region for the global fit, the valence quark CSV 
PDFs could be either four times as large as that predicted by Sather 
or Rodionov, or it could be three times as big with the opposite sign.  
The great value of the MRST global fit is that CSV distributions with 
this shape, and with values of $\kappa$ within this range, will not 
disagree seriously with any of the high energy data used to 
extract quark and gluon distribution functions. 

The MRST group also searched for the presence of charge 
symmetry violation in the sea quark sector.  Again, they chose a 
specific form for sea quark CSV, dependent on a single parameter, \IE   
\bea 
\bar{u}^n(x) &=& \bar{d}^p(x)\left[ 1 + \delta \right] \nonumber \\  
\bar{d}^n(x) &=& \bar{u}^p(x)\left[ 1 - \delta \right] 
\label{eq:seaCSV}
\eea  
With the form chosen, the net momentum carried by antiquarks is 
approximately conserved; violation of momentum conservation was found to be 
very small in the kinematic region of interest.  

Somewhat surprisingly, evidence for sea quark CSV in the global fit 
is substantially stronger than that for valence quark CSV.  As shown in 
the top curve in Fig.\ \ref{Fig:chi2}, the best fit is obtained for 
$\delta = 0.08$, corresponding to 
an 8\% violation of charge symmetry in the nucleon 
sea. The $\chi^2$ corresponding to this value is substantially 
better than with no charge symmetry violation, primarily because of 
the improvement in the fit to the NMC $\mu-D$ DIS data \cite{Ama91,Arn97} 
when $\bar{u}^n$ is increased.  The fit to the E605 Drell-Yan data 
\cite{E605} is also substantially improved by the sea quark 
CSV term.  

\begin{figure}
\includegraphics[width=2.5in]{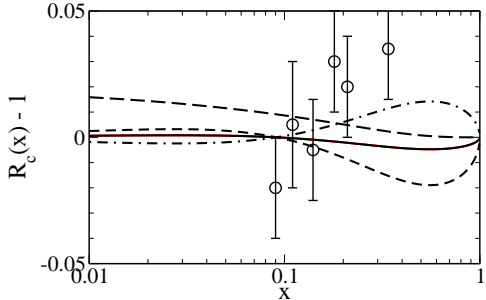}
\caption{The ``charge ratio'' $R_c -1$ defined in Eq.\ 
\protect{\ref{eq:Rc}}, compared with charge symmetry violating PDFs 
obtained by the MRST global fit, Ref.\ \protect{\cite{MRST03}}.  
Data is obtained from CCFR neutrino cross sections of  
Ref.\ \protect{\cite{Sel97}} and NMC muon DIS, 
Ref.\ \protect{\cite{Ama91,Arn97}}. Long-dashed curve: sea quark 
CSV with best fit parameter $\delta = 0.08$ in Eq.\ 
\protect\ref{eq:seaCSV}; solid curve: valence quark CSV corresponding to best 
fit value $\kappa = -0.2$ in Eq.\ \protect\ref{eq:CSVmrst}; short 
dashed curve: valence quark CSV for $\kappa = -0.8$; dash-dot curve: valence 
quark CSV for $\kappa = +0.65$. 
\label{Fig:MRSTRc}}
\end{figure}

We can check the MRST results by comparing them with the charge 
ratio measurements that were summarized in Sect.\ \ref{Sec:Csymmexp}.  
Eq.\ \ref{eq:partCSV} relates the CSV parton distributions to the 
charge ratio through 
\beast
 R_c - 1 &=& {3x \over 10 Q(x)}\left( \delta u(x) + \delta \bar{u}(x) - 
 \delta d(x) - \delta \bar{d}(x) \right) \, . 
\eeast
We have taken the valence and sea quark PDFs from MRST 
\cite{Thopc}, and 
in Fig.\ \ref{Fig:MRSTRc} we plot them against the experimental 
charge ratio obtained using the CCFR charged-current neutrino structure 
function and the NMC muon structure function.  The long-dashed curve 
corresponds to the sea quark CSV term with the best value 
$\delta = 0.08$ from the MRST fit.  The remaining curves correspond to 
various values for valence quark CSV.  The solid curve corresponds to 
the best value $\kappa = -0.2$, the short-dashed curve corresponds to 
$\kappa = -0.8$, and the dash-dot curve corresponds to $\kappa = +0.65$.  
The latter two correspond to values of $\kappa$ at the 
90\% confidence level for the MRST fit to valence quark CSV.  
For values $x < 0.085$, we have not included the charge ratio 
experimental data since these changed significantly in the CCFR re-analysis.  
However, the re-analyzed data resulted in a 
charge ratio that agreed with unity at about the 2-3\% level, even down to 
$x \approx 0.01$.  

Both the valence and sea quark CSV distributions of MRST are in good 
agreement with the experimental limits from the CCFR/NMC data.  At 
every $x$ value the phenomenological values are within two standard 
deviations of the data.  This is understandable, since both the NMC 
and CCFR data were included in the MRST global fit that extracted the 
CSV parameters.  We have not included overall normalization errors, of 
roughly 2\%, on both the CCFR and NMC data.  Since the best fit 
valence quark PDFs of MRST (the solid curve in Fig.\ \ref{Fig:MRSTRc}) 
are very 
close to the theoretical predictions of Sather and Rodionov, the  
theoretical predictions are well within the upper limits on parton 
CSV from the best experiments. CSV effects in reasonable agreement 
with high energy data are substantially larger than predicted by 
theory; valence CSV effects could be four times as large as predicted 
by Sather or Rodionov (or three times as large with the opposite 
sign), and sea quark CSV effects are also significantly larger 
than theoretical predictions \cite{Ben98}.   

The only remaining issue is whether the theoretical and phenomenological 
PDFs agree with the limits on parton momentum, estimated by MRST 
as relative errors of about 4\%.  The best value obtained by 
MRST was $\kappa = -0.2$.  With this value, one obtains the 
relative momentum change for valence neutrons 
\be
 {2|\delta U_{\V}| \over U_{\V} + D_{\V}} = 
 {2\delta D_{\V} \over U_{\V} + D_{\V}} \approx 1\% 
\label{eq:CSVmom} 
\ee 
This value is well within experimental limits.  Even the 
largest value of $\kappa$ within the 90\% confidence limit, $\kappa = -0.8$, 
corresponds to a momentum change of 4\%.  Thus all of the phenomenological 
valence CSV distributions (and the theoretical distributions of 
Sather and Rodionov {\it et al.}) correspond to momentum values within 
experimental limits.   

\section{Charge Symmetry Violation and Neutrino DIS 
Reactions\label{Sec:CSVPW}}

In 1973, Paschos and Wolfenstein \cite{Pas73} suggested that 
the ratio of neutral-current and charge-changing neutrino 
cross sections on isoscalar targets could provide an independent 
measurement of the Weinberg angle ($\sintW$).  The Paschos-Wolfenstein 
(PW) ratio $R^-$ is given by 
\be
 R^- \equiv { \snuNC - \snubNC \over 
 \rz \left(\snuCC - \snubCC \right)} = {1\over 2} - \sintW  .
\label{eq:PasW} 
\ee  
In Eq.\ \ref{eq:PasW}, $\snuNC$ and $\snuCC$ are respectively the 
neutral-current and charged-current inclusive, total 
cross sections for neutrinos (or antineutrinos) on an 
isoscalar target.  The quantity $\rho_0 \equiv M_{\W}/(M_{\Z}\,\cos 
\theta_{\W})$ is one in the Standard Model. Not only is the 
PW ratio independent of parton distribution functions, 
but in this ratio a large number of partonic corrections either cancel 
or are minimized.   

The NuTeV group has measured neutrino charged-current and 
neutral-current cross sections on iron \cite{NuTeV} and  
extracted a value for $\sintW $ equal to 
$0.2277 \pm 0.0013 ~(stat) \pm 0.0009 ~(syst)$.  This 
value is three standard deviations above the measured fit to 
other electroweak processes, $\sintW = 0.2227 \pm 0.00037$ \cite{EM00}.  
Such an effect can be interpreted as a 1.2\% decrease in the 
left-handed coupling of light valence quarks to the weak neutral 
current.  In addition to the Weinberg angle, several of the 
parameters of the Standard Model are constrained by the very precise 
measurements at electron-positron colliders \cite{EM00}. Indeed, many of these 
parameters are determined at about the 0.1\% level.  Consequently, the 
discrepancy between NuTeV and electromagnetic measurements of the 
Weinberg angle is surprisingly large.  It suggests evidence of physics 
beyond the Standard Model, although Davidson \EA \cite{Davidson:2001ji} 
have shown that it is quite difficult to produce ``new physics'' (\IE 
beyond the Standard Model) that fits the NuTeV experiment without violating 
other experimental constraints. In this paper, we will examine  
isospin-violating corrections to the NuTeV experiment. 

The NuTeV group did not directly measure the PW ratio, but instead 
measured the neutral-current to charged-current ratios $\Rnu$ and 
$\Rnub$.  These quantities, and their relation to the PW ratio, are given 
by 
\bea 
 \Rnu &\equiv& {\snuNC \over \rz \snuCC}, \hspace{0.25truein}  
 \Rnub \equiv {\snubNC \over \rz \snubCC} \nonumber \\ 
 R^- &=& {\left( \Rnu - \Rnub \right) \over 1 - r\, \Rnu} ~,  
 \hspace{0.25truein} r = {\snubCC \over \snuCC} 
\label{eq:Rnu}
\eea 
After measuring the ratios $\Rnu$ and $\Rnub$, the NuTeV group obtain 
the Weinberg angle through a detailed Monte Carlo simulation of the 
experiment.  As a result, it is substantially more 
difficult to estimate the errors due to corrections to this experiment. 
For partonic corrections, the NuTeV group \cite{NuTeV2}
have provided functionals that 
estimate the change in an experimental quantity ${\cal E}$ due to changes 
in a parton distribution $\delta(x)$:
\be
  \Delta {\cal E} \equiv \int_0^1 \, 
  F\left[ {\cal E}, \delta; x\right] \, \delta(x) \, dx .
\label{eq:Func}
\ee
In Eq.\ \ref{eq:Func}, $F\left[ {\cal E}, \delta; x\right]$ is the 
functional, and the net change in the observable is obtained by 
integrating over Bjorken $x$. The NuTeV collaboration provided functionals 
appropriate for both valence and sea quark CSV.    

The correction to the Paschos-Wolfenstein ratio arising from isospin 
violation in the parton distribution functions has the form  
\bea 
 \Delta R_{\CSV} &=& \left[ 1 - {7\over 3}\sintW + {4\alpha_s \over 9\pi}
 \left( {1\over 2} - \sintW \right) \right] \nonumber \\ 
 &\cdot& {\delta U_{\V} - 
 \delta D_{\V}  \over 2(U_{\V} + D_{\V}) } \, , \nonumber \\  
 {\rm where} \hspace{0.6cm} \delta Q_{\V} &=& \int_0^1 \, 
  x\,\delta q_{\V}(x) \,dx \, . 
\label{eq:CSV} 
\eea
Note that the isospin violating correction to the PW ratio arises 
entirely from charge symmetry violation in the valence PDFs.  Inserting the 
quark model results of 
Sather \cite{Sat92} and Rodionov {\it et al.}~\cite{Rodionov:cg} into Eq.\ \ref{eq:CSV}, 
the CSV correction to the PW ratio is found to be 
$\Delta R_{\CSV} = -0.0021$ using Sather's parton distributions, and 
$\Delta R_{\CSV} = -0.0020$ using the CSV distributions of Rodionov 
\EA \cite{Lon03}.  These are theoretical {\it predictions}, as the Sather 
and Rodionov papers were published in 1992 and 1994, 
respectively, well before the analysis of the NuTeV data.  Like the 
isoscalar corrections 
to the PW ratio, the CSV corrections involve the ratio of identical 
moments of parton distributions, and are thus independent of $Q^2$; 
consequently they do not depend on any details of QCD evolution.  

The negative sign of the result means that CSV corrections will decrease 
the discrepancy between the NuTeV result for the Weinberg angle, and 
the Weinberg angle extracted from the extremely precise data obtained 
from electron-positron colliders near the $Z$ mass.  
Indeed, the predicted CSV corrections 
to the PW ratio remove roughly 40\% of the magnitude of the 
NuTeV anomaly.  As discussed in the preceding Section, less than 1\% of 
the valence quark momentum is carried by the 
relevant CSV combination, so this is well within the 
experimental limits on quark momentum.  

Londergan and Thomas \cite{Lon03b} showed that Sather's formula of  
Eq.\ \ref{eq:Satanl} provided an analytic estimate of CSV 
corrections to the PW ratio; taking the second moment of Sather's 
equations gives    
\be 
 \delta D_{\V} = \frac{\delta M}{M} D_{\V} + \frac{\delta m}{M} \,; 
  \hspace{0.6cm}  
 \delta U_{\V} = \frac{\delta M}{M} \left( U_{\V} - 2 \right) 
\label{eq:intUv}
\ee  
Sather's approximation predicts that 
the CSV correction to the PW ratio depends only on 
the fraction of nucleon momentum carried by up and down valence quarks.  
Eq.\ \ref{eq:intUv} predicts that $\delta D_{\V}$ will be 
positive and $\delta U_{\V}$ negative, in agreement with the  
theoretical model results.  It is especially interesting that 
Sather's analytic approximation allows one to 
calculate the CSV correction to the PW ratio directly from valence quark 
PDFs, without ever having to calculate specific CSV distributions.  
Using Eq.\ \ref{eq:intUv}, we also calculated the CSV correction to 
the PW ratio using the CTEQ4LQ phenomenological parton 
distribution \cite{CTEQ4} at $Q^2 = 0.49$ GeV$^2$, obtaining  
$\Delta R_{\CSV} = -0.0021$ -- again in agreement with the results 
obtained from the two quark model calculations. Eq.\ \ref{eq:intUv} 
shows why the CSV corrections are so similar in various models;  
the correction depends only on the momentum carried by up and down 
valence quarks, a quantity well determined in PDFs.    
 
Since the NuTeV group \cite{NuTeV,NuTeV2} did not directly 
measure the PW ratio, one must multiply the CSV PDFs with 
the relevant functional in Eq.\ \ref{eq:Func} \cite{NuTeV2}.  
This requires evolving the CSV parton distribution 
functions from the quark-model scale to $Q^2 = 20$ GeV$^2$, the 
average momentum transfer for the NuTeV experiment. After QCD evolution 
and folding with the NuTeV functionals, the resulting CSV corrections 
to the NuTeV result are $\Delta R_{\CSV} = -0.0015, -.0017$ and 
$-0.0014$ for the Rodionov, Sather and CTEQ4LQ PDFs, respectively.  
The CSV corrections decrease the NuTeV discrepancy in the Weinberg 
angle by about 30\%, or one standard deviation.  
While some groups have 
obtained substantially smaller estimates for the charge symmetry 
violating contribution to the NuTeV anomaly \cite{NuTeV2,Cao:2003ny}, 
in the first case the constraint on baryon number was not respected 
while the second paper assumed a completely different mechanism for 
parton CSV.  On the other hand, as we discuss next, the phenomenological 
values for CSV allow considerably larger isospin violation than 
any of the theoretical estimates. 

Valence quark CSV makes a substantially larger 
contribution than sea quark CSV to the extraction of the Weinberg angle from 
neutrino DIS.  Using the sea-quark CSV and the best-fit value for valence 
quark CSV obtained by the MRST group, would remove roughly 
$1/3$ of the NuTeV anomaly.  The value $\kappa = -0.6$, within the 
90\% confidence limit found by MRST, would completely remove the 
NuTeV anomaly, while the value $\kappa = +0.6$ would double the 
discrepancy.  The MRST results show that isospin 
violating PDFs are able to completely remove the NuTeV anomaly in the 
Weinberg angle, or to make it twice as large, without serious disagreement  
with any of the data used to extract quark and gluon PDFs. 

There are other possible corrections to the Weinberg angle extracted 
by the NuTeV experiment, but for the most part the corrections are 
small and/or well understood.  The correction 
for the neutron excess in iron is large, but apparently well known 
\cite{Davidson:2001ji,NuTeV2}.  Radiative corrections are also large, 
but also believed to be under control \cite{Bar79}, although the 
radiative corrections are being re-analyzed \cite{McF04}.  The situation 
with strange quarks is not certain. Spontaneously broken chiral SU(3) $\times$ SU(3)
symmetry implies an asymmetry between $s(x)$ and $\bar s(x)$~\cite{Signal:1987gz}. 
As a direct consequence of such an asymmetry there is a correction to the 
extraction of the Weinberg angle that depends on the momentum asymmetry 
\be
  S_{\V} = \int_0^1 \,x\left[ s(x)-\bar{s}(x) \right] \,dx  \, . 
\label{eq:Sasym}
\ee 
A positive (negative) value for $S_{\V}$, meaning that strange quarks 
carry more (less) net momentum than strange antiquarks, would decrease    
(increase) the size of the Weinberg anomaly. The strange quark and 
antiquark distributions can be extracted from 
the cross sections for opposite sign dimuons in reactions induced by 
neutrinos or antineutrinos; such reactions have been measured by 
the CCFR \cite{Baz93} and NuTeV \cite{Gon01} collaborations. The 
NuTeV group has extracted strange and antistrange PDFs from the 
dimuon production reactions; they obtain values for 
$S_{\V}$ that are consistent with zero, or perhaps small and negative 
\cite{NuTeV2}.  The CTEQ group on the other hand \cite{Kret04,Oln03,Kret03}, 
has included the dimuon data in a global fit of PDFs; they obtain positive 
values for $S_{\V}$ that could account for roughly one-third of the 
Weinberg angle anomaly. In both NuTeV and CTEQ analyses  
the strange quark distributions are determined from the same CCFR/NuTeV 
dimuon data, but at present the two analyses obtain qualitatively 
different results for the strange quark momentum asymmetry.    

In conclusion, it seems that the magnitude of 
CSV effects allowed by the MRST fit makes charge symmetry 
violation one of the only viable explanations for the anomalous value of 
the Weinberg angle obtained in the NuTeV neutrino experiment. If  
CSV effects are sufficiently large to remove the Weinberg angle 
anomaly, such effects should be visible in various other experiments.  
We had previously suggested several experiments that could potentially 
reveal the presence of parton CSV \cite{Lon98}. However, the 
magnitude of those effects were based on theoretical calculations that  
predicted substantially smaller CSV effects than are  
allowed by MRST. It is clearly of great interest now to investigate this 
issue experimentally.  

\section*{Acknowledgments\label{Sec:Acknwl}}

This work was supported in part [JTL] by National
Science Foundation research contract PHY0302248 and by [AWT] 
DOE contract DE-AC05-84ER40150, under which SURA operates 
Jefferson Laboratory. The authors wish to thank G.P. Zeller and K. 
McFarland of the NuTeV collaboration, W. Melnitchouk at JLab, and R.S. 
Thorne from the MRST group, for useful discussions regarding the issues
presented here.


\begin{thebibliography}{99}
\bibitem{Shadnt} Nuclear effects in neutrino reactions should consist 
of ``shadowing'' corrections at small $x$, ``EMC''-type corrections 
at intermediate $x$, and Fermi-motion effects at large $x$.  Fermi-motion 
effects are expected to be identical to those seen in 
muon DIS; shadowing effects for neutrinos should differ from 
muon DIS on the same nuclear target, see Ref.\  \protect\cite{Bor98b}; 
and EMC effects may also differ between muons and neutrinos.
\bibitem{Shad} In the published values of the charge ratio, the 
discrepancy at small $x$ is even larger than shown here. The 
calculations shown here included estimates of shadowing for neutrino CC 
reactions 
that differed from the shadowing observed in DIS for charged leptons, 
see Ref.\ \protect\cite{Bor98b}.
\end{thebibliography}
\end{document}